\documentclass[doublecolumn,10pt]{IEEEtran}
\usepackage[lined,boxed,commentsnumbered,ruled,linesnumbered]{algorithm2e}
\usepackage{cite}
\usepackage{caption}
\usepackage{subcaption}
\usepackage{graphicx, amsmath, amssymb, caption, url, authblk,lineno, diagbox}
\usepackage[bottom]{footmisc}

\graphicspath{{../figure/}, {./figure/}}

\usepackage{color}
\newcommand{\tabincell}[2]{\begin{tabular}{@{}#1@{}}#2\end{tabular}}

\newcommand{\beq}{\begin{equation}}
\newcommand{\eeq}{\end{equation}}

\usepackage{amssymb} 
\usepackage{pifont}
\newcommand{\cmark}{\ding{51}}%
\newcommand{\xmark}{\ding{55}}%
\usepackage{amsthm} 

\begin{document}

\title{Small-floating Target Detection in Sea Clutter via Visual Feature Classifying in the Time-Doppler Spectra}
\author{Yi Zhou, Yin Cui, Xiaoke Xu, Jidong Suo, Xiaoming Liu \thanks{
Yi Zhou, Jidong Suo and  Xiaoming Liu are with the Department of Electronic Information Engineering, Dalian Maritime University, Dalian, 116026, China. Email: (\{yi.zhou, sjddmu, lxmdmu\}@dlmu.edu.cn).  Yin Cui is with Dalian Shipbuilding Industry Design and Research Institute, Dalian, 116005, China. Email:(cui\_ying@dsic-design.cn). Xiaoke Xu is with College of Information and Communication Engineering, Dalian Minzu University, Dalian 116600, China. Email: xuxiaoke@foxmail.com.
}}
{}

\maketitle

\begin{abstract}
It is challenging to detect small-floating object in the sea clutter for a surface radar. In this paper, we have observed that the backscatters from the target brake the continuity of the underlying motion of the sea surface in the time-Doppler spectra (TDS) images. Following this visual clue, we exploit the local binary pattern (LBP) to measure the variations of texture in the TDS images. It is shown that the radar returns containing target and those only having clutter are separable in the feature space of LBP. An unsupervised one-class support vector machine (SVM) is then utilized to detect the deviation of the LBP histogram of the clutter. The  outiler of the detector is classified as the target. In the real-life IPIX radar data sets, our visual feature based detector shows favorable detection rate compared to other three existing approaches.
\end{abstract}


%
\IEEEpeerreviewmaketitle

\section{Introduction}
Detecting small-floating object in sea clutter is a challenging task for marine surveillance radar. Since the amplitude of the backscatters of the  clutter are target-like in the low-grazing viewing aspect, it is hard to separate small target from the clutter directly on the amplitude of  radar returns \cite{WattsProc90}. Coherent signal processing may provide help by measuring the Doppler shift of the returns, if the target has enough radial velocity. However for the surface-floating object, their slow-speed motion is difficult to be distinguished from the spreading of the sea waves \cite{HaykinProc02}. 

Recently, feature based detectors show the great effect on the small-floating object detection \cite{XuTAP10,ShuiTAES14,ShiTGRS18} in the real-life IPIX \cite{ipixDataset} radar data sets. The key of these detectors is to find new feature space which can easily separate the target and clutters. In \cite{XuTAP10}, fractal statistics of the amplitude of the returns are proposed to capture the fractal differences between sea clutter and the target. In \cite{ShuiTAES14} three features of the sequential returns: the relative amplitude, relative Doppler peak height, and relative entropy of the Doppler amplitude spectrum are jointly combined to distinguish the target from sea clutter. 
In \cite{ShiTGRS18} normalized time frequency distribution (NTFD) on the 2D image of Time Doppler Spectra (TDS) is proposed to enhance the visual discriminability between the returns of clutters and of targets. Other three heuristic features: accumulation of the maximum of the time slices on NTFD,  the number and maximum size of the connected regions on the binary image of the threshold NTFD are combined to model the 3D-feature space.  Like in \cite{ShuiTAES14}, a convex hull based one-class classifier is used to detect the feature of the target cell, which is far away to the inner convex hull of the 3D feature of the clutter-only cells. In \cite{XuTAP10, ShuiTAES14, ShiTGRS18}, they all joints multiple features to improve the detection rate. 
\textbf{The choice of the multiple kinds of features are often based on the experimental observations, not on the theoretically analysis. It reflects that the discriminable feature is extremely valuable and  hard to be found.}

\begin{figure}
     \centering
     \begin{subfigure}[b]{.5\textwidth}
         \centering
         \includegraphics[width=\textwidth]{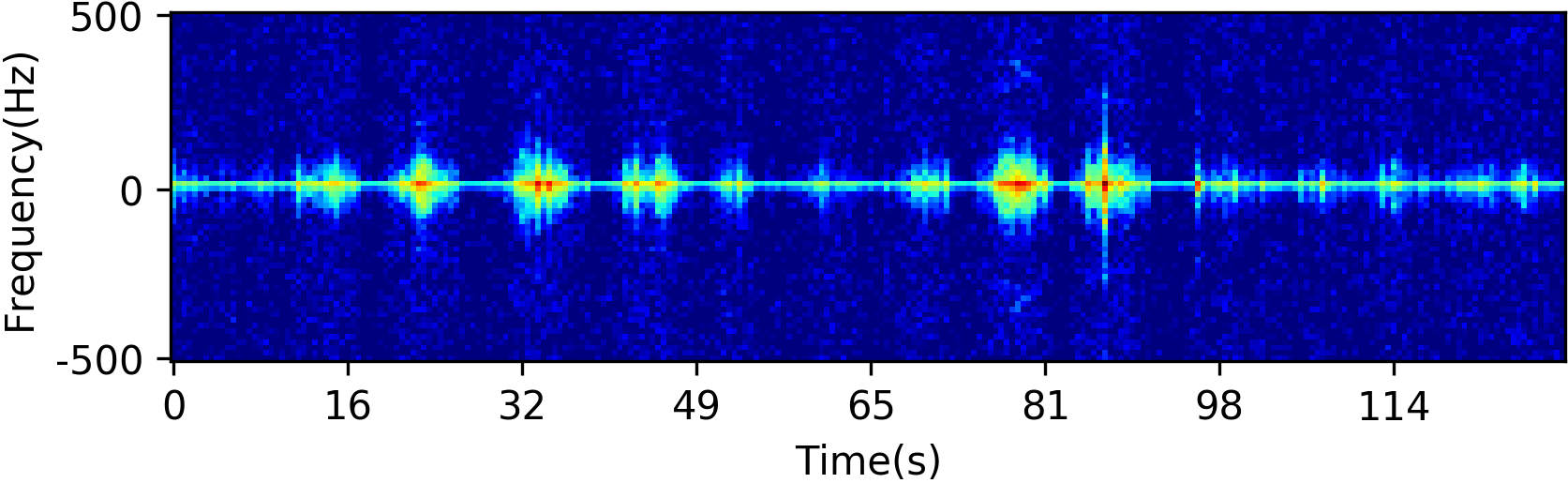}
         \caption{clutter-only image has constrained intensities in the low frequency band (around horizontal zero-frequency line).}
         \label{fig.tds.clutter}
     \end{subfigure}
     \vfill
     \begin{subfigure}[b]{.5\textwidth}
         \centering
         \includegraphics[width=\textwidth]{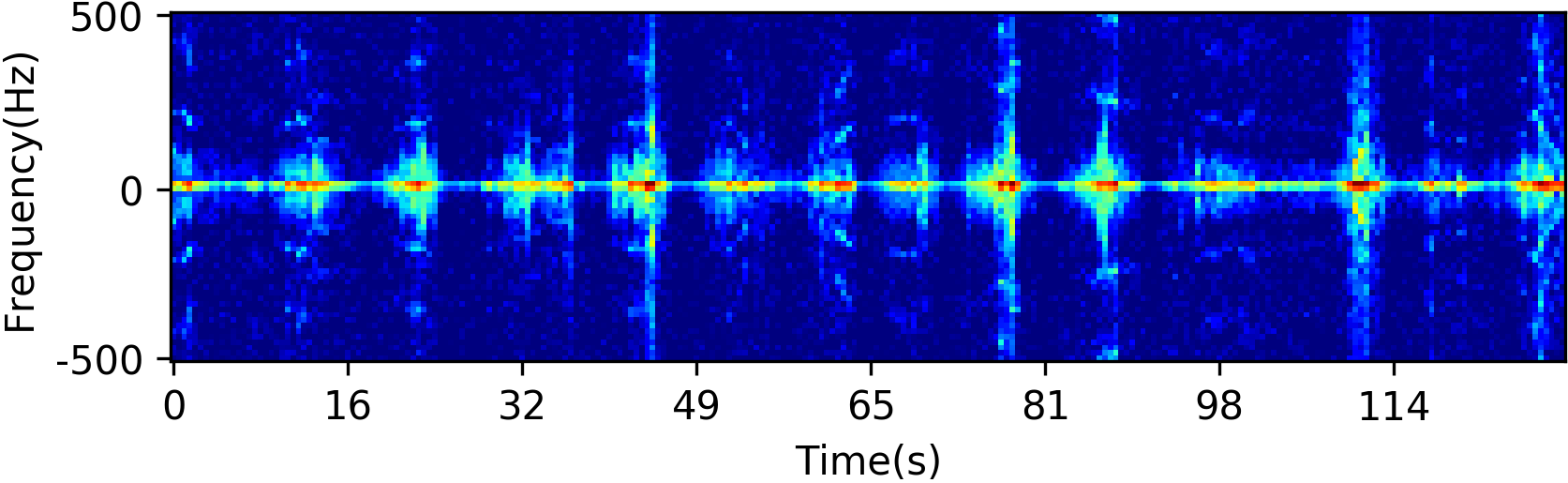}
         \caption{target-contained image shows the energy spreading in the whole frequency band.}
         \label{fig.tds.target}
     \end{subfigure}
    \caption{Time-Doppler Spectra (TDS) image comparisons between (a) the clutter-only cell, which has constrained energy near the zero-frequency band, and (b) the target-contained cell. It is shown that in the target-contained cell, the underlying Doppler frequency  of the waves in the low frequency band is interfered by the target. 
Furthermore, this interference has caused new medium and high frequency components, which looks like speckles in the target-contained TDS image. 
}
    \label{fig.tds.comparison}
\end{figure}

It is noted that TDS image are often \textbf{transformed} to produce specifically numerical feature, such as entropy in \cite{ShuiTAES14, LiGRSL19}, region size and accumulation in \cite{ShiTGRS18}. These features are \textbf{customized} in low dimensions for the sake of the convenience of training.
Is there any \textbf{generic} visual feature available to characterize the target and clutter in the \textbf{original} TDS?
In Fig.\ref{fig.tds.comparison}, it shows two types of TPS images. The upper one is the spectra only has clutter, while the one below gets both the floating target and clutter. 
\textbf{It is shown that target-contained image has more spreading energy in the frequency axis, which has caused distinct texture variation visually in TDS image.}


In image processing and computer vision community,  the description of image texture affects the performance of the object detection, recognition and scene understanding. The study of image texture has long-travelled efforts. Among them, the local binary patterns (LBP) is the time-tested feature \cite{PietikBook2011}. It was first proposed to transform the local-intensity differences to the ordered signs and represent them as binary number in \cite{OjalaICPR94}.  In \cite{OjalaTPAMI02}, LBP was further extended to the rotation-variant `uniform', where it is tolerated to monotonic illumination changing and image rotation. Latter, LBP feature was applied to facial recognition in time-spatial domain \cite{ZhaoTPAMI07}.  LBP was also validated to detect the tumor in the X-ray images \cite{LinderDP2012}. 
In the high resolution radar image, modified LBP is applied to classify ground echos in the plan position indicator (PPI) image \cite{HedirSIVA18}, or to match patches in the synthetic aperture radar (SAR) image \cite{GhannadiGRSL19}. 


Inspired by visual difference of the texture between the two kinds of TDS images in Fig.\ref{fig.tds.comparison}, we propose LBP to model the distinguishable feature spaces for separating the target-contained TDS from the clutter-only ones. To handle the high dimensions of the texture, we resort to the one-class SVM based classifier \cite{ScholkopfNC01}, which is proved to be effect to find the outlier among the imbalanced and unlabelled samples. 



The main contributions of this paper are two-fold:

1. Use the generic visual texture feature LBP to characterize the sea clutter and floating target in the TDS image. It demonstrates that the appearance of target changes the inherent properties of the clutter in the LBP feature space.  

2. Explore the one-class SVM classifier to detect the sudden variations of feature in a group of TDS images. To balance the conflict between the high dimensions of the feature and the limited training samples, we prove that trained boundary with the imbalanced samples is more closer to the outlier. Therefore,
we sort the distances of all samples to the decision boundary and find their minimum as the outlier -- the floating target.  

The remainder of this paper is organized as follows. Section \ref{sec.lbp} introduces the LBP histogram to characterize the TDS image in the real radar data sets. Section \ref{sec.svm} designs the floating-target detection in the framework of one-class SVM and analyses the theory basis. Section \ref{sec.exp} discusses the  experimental results on the IPIX data sets. Section \ref{sec.con} concludes this work.


\section{LBP properties of sea clutter}
\label{sec.lbp}
\subsection{IPIX radar data sets}
The study of detection of small-floating object in sea clutter relies heavily on the real-life radar data sets. IPIX\cite{ipixDataset} is the most famous and widely used real radar data for detecting small targets in  sea clutter in the radar research community. In Nov. 1993, a group led by Simon Haykin used the X-band (9.39 GHz in frequency and about 3cm in wavelength) IPIX radar on a clifftop near Dartmouth, 
in the east coast of Canada, to  collect high-resolution, coherent, and polarimetric radar returns.  The radar can work in the staring mode -- transmitted and received  pulses only in one azimuth angle, where a testing target was anchored in that direction around 2.6 km far away. Tested target was a ball with 1 meter diameter, which was made by Styrofoam and  wrapped with wire mesh. Each data set collected coherent data in like-polarized (VV, HH) and cross-polarized (VH, HV) configuration in $2^{17}$ complex numbers (around 131 seconds) for 14 range cells. 
In 1998, the IPIX radar with upgraded data precision was operated in Grimsby, on the shore of Lake Ontario. Collected data sets got a small floating boat in 28 adjacent range cells in one minutes ($6\times10^{4}$ complex numbers for each range cell).
In IPIX data sets, the cell contained the target was called the primary cell, the neighbour cells affected by  the target is termed as the secondary cells, and the remaining are clutter-only cells. 

\subsection{Time Doppler Spectra}
\label{subsec.tds}

The coherent radar such as IPIX receives the signal in complex form which contains both the inphase ($I$) and quadrature ($Q$) part. It provides the availability to measure both  the amplitude and the phase of the returns \cite{HaykinProc02}. Motion of the clutter relative to the radar site would bring a pulse-to-pulse change in the phase, which is termed as Doppler frequency shift. This shift $f_d$ is defined as :
\beq
f_d = \frac{2v}{\lambda},
\eeq
where $\lambda$ is the wavelength and $v$ is the radial velocity between the radar and the moving object. The spread of the Doppler frequency indicates the evolution of the motion of the backscatters, which could be caused either by the floating target or the moving waves.  And the energy of this frequency would be captured by doing the windowed short-time  Fourier transforming on the sequential complex radar returns.

It is noted that there are continuous waves of various heights, lengths and directions on the sea surface \cite{ipixDataset}. 
Observing the Doppler spectrum in a long time (one or two minutes), this continuum would be reflected as a periodical distribution in the Doppler frequency. Record the Doppler spectrum versus time, it will form a 2D image like Fig.\ref{fig.tds.comparison}, we call it time Doppler spectra (TDS) in this paper.


Here we take the IPIX data set for example to show how to compute a TDS image.
Staring in one beam position, IPIX radar data stored $2^{17}$ and 60000 orthogonal $I$ and $Q$ samples for each range cell  in the 1993 and 1998 data sets respectively. 
Divide the long sequential echo into  non-overlapping segments, each segment has  the length of $l$.  Suppose $w$ segments are obtained. 
For each segment, implementing the windowed short-time Fourier transformation gets $l$-length Doppler spectrum. Re-scale it to a shorter $h$-length  spectrum.  Stack all the $w$ Doppler spectra vertically, the 2D time-Doppler spectra ($h \times w$) is acquired for each range cell.  
%

Fig.\ref{fig.tds.comparison} illustrates two TDS images which are extracted from the file indexed 135603 
in 1993 data sets\cite{ipixDataset}. One for the clutter only cell and another for the primary cell which contains the small target. Compared to TDS images of the clutter only, target contained TDS image shows the energy spread in the medium and high frequencies and clearly tells that the appearance of target affect the structure of the clutter in the Doppler frequency domain. 
This is explained in the chapter 2 of \cite{WardBook06} that within a radar resolution cell, backscatters from different small surfaces, which move relative to each other, may cause interference in the amplitude of the returns. In Fig.\ref{fig.tds.comparison}, we have observed this interference in the \textbf{time-frequency domain}, where the texture of the two types of images are different. Therefore we can directly model the clutters based on their texture information in the TDS image and quantize the variations on texture to magnify the interference.

\subsection{LBP features}

\begin{figure}[htb]
    \includegraphics[width=\linewidth]{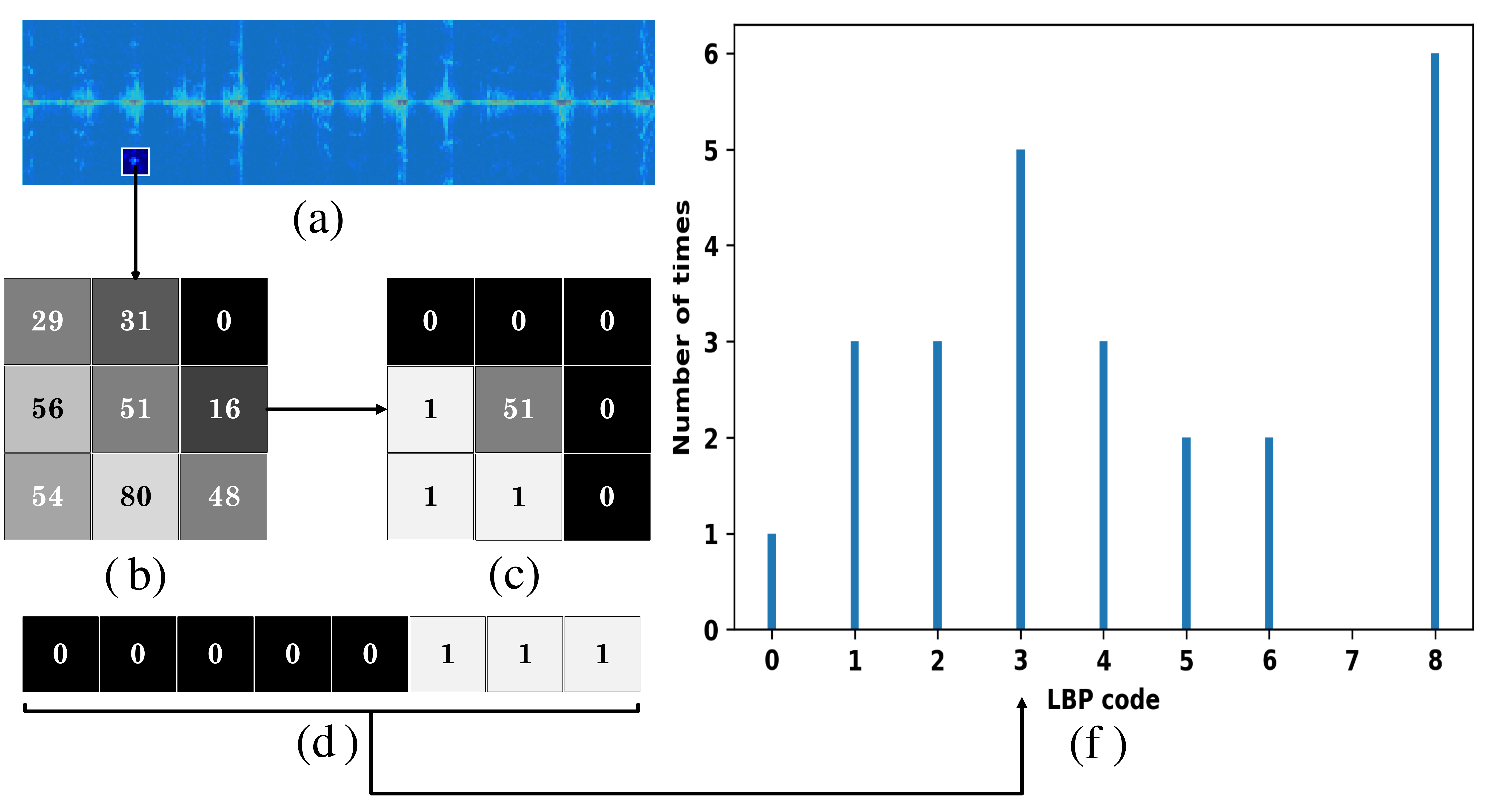}
    \caption{Demonstrate the procedure of computing LBP histogram. (a) Mark a grid region of a target-contained TDS image, which is a typical speckle in the medium frequency. (b) Show the intensities of the center pixel of the grid and of its 8 neighbours. (c) Neighbours with higher intensities than the central are labeled 0, otherwise 1. (d) Convert the ordered binary string into the `uniform' LBP code 3. (f) Count all the LBP codes for each pixel in the grid and obtain the normalized LBP histogram for the marked region.}
    \label{fig.demo.lbp}
\end{figure}

In LBP,  the characteristic of local texture is approximated by the joint difference distribution\cite{OjalaTPAMI02} as follows:
\beq
T \approx t\left( g_0 -g_c, g_1-g_c, ..., g_{P-1}-g_c\right).
\eeq
Here $g_c$ means the gray scale of the central pixel, $g_0$, ..., $g_{P-1}$ are the intensities of the $P$ circular neighbor pixels around $g_c$. $T$ computes the occurrences of different patterns in local region by a $P$-dimensional histogram. 
In the flat regions, the differences in all directions are zero. For the sloped edge, highest difference occurs in the gradient direction and zero values appear along the edge. For a spot (the single lighting or dark pixel),  $T$ holds omnidirectional differences. 
To further generalize $T$ to resist the varying of the illumination, signs of the differences take the place of their exact values as follows:
\beq
T \approx t\left( s(g_0 -g_c), s(g_1-g_c), ..., s(g_{P-1}-g_c)\right),
\eeq
where
\beq
s(x) = \left\{\begin{matrix}
1 & x\geqslant 0\\ 
0 & x<0
\end{matrix}\quad .\right.
\eeq
Here, the $P$ denotes the number of pixels around central pixel $c$.  
If the pixels get lower intensity than the value of $g_c$, their signs are set 0.  Otherwise, they are set 1. 
To reduce the dimensions of the distribution, the $P$-dimension inputs of  $t(\cdot)$ are simplified as an ordered binary code in:
\beq
\label{eq.lbp.code}
LBP_{P,R}=\sum_{p=0}^{P-1}s\left(g_p-g_c\right)2^p.
\eeq
Here $R$ represents the radius of $P$ circular neighbor pixels. 
Multiplying a binomial factor $2^p$ for each sign in order and do accumulating in \eqref{eq.lbp.code}, one kind of spatial structures in the local texture is simplified as a definite number, termed  the $LBP$ code. So, the local texture is now modeled as the distribution of the 'uniform' LBP codes, which is also called the LBP histogram in the numerical implementation.
It is observed that majority of the codes have at most two transitions between 1 and 0 in a circle \cite{OjalaTPAMI02}. These codes are termed as ‘uniform’ patterns and their maximum value only relates to the number of the neighbor pixels $P$.

In Fig.\ref{fig.demo.lbp}, we demonstrate the procedure of computing LBP histogram for a region of interest (ROI) in one target-contained TDS image. In Fig.\ref{fig.demo.lbp} (a) the white rectangle marks the region, where the LBP histogram is wanted. Then in Fig.\ref{fig.demo.lbp} (b), one central pixel with 51 intensity and its 8 neighbour  pixels with different intensities are shown. In Fig.\ref{fig.demo.lbp} (c) the neighbour pixels is compared to the central 51 and get the `0, 1' string, which is converted to the binary code according to the \eqref{eq.lbp.code} in the Fig.\ref{fig.demo.lbp} (d). Three continuous `1's equal to `3' of the `uniform' LBP code. Following this way, all the pixels in the ROI obtain their own LBP codes. The LBP code distribution of the ROI is modeled by the LBP histogram in  Fig.\ref{fig.demo.lbp} (f).

\subsection{Separability of the target and sea clutter in the LBP feature space}
\label{subsec.LBPsepa}

\begin{figure}[!t]
    \includegraphics[width=\linewidth]{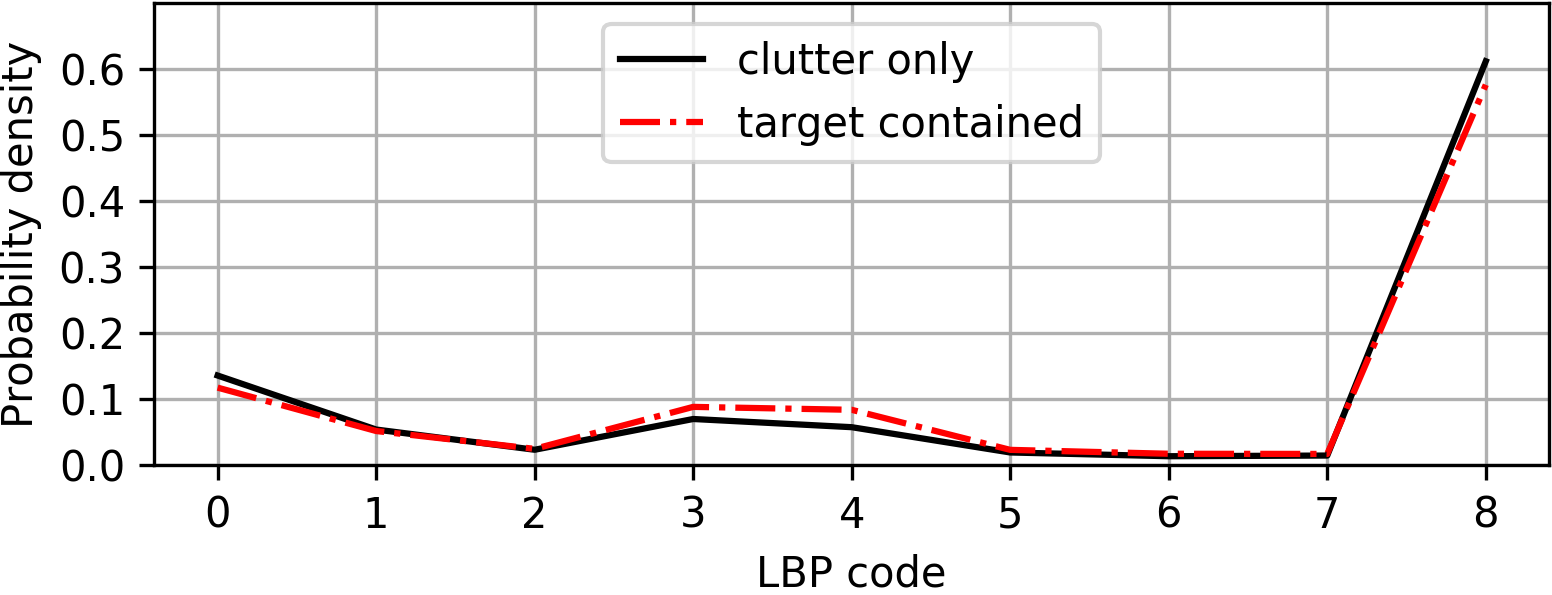}%
    \caption{LBP histogram comparison in the `HH' mode of the file indexed 135603
    in Dartmouth(1993) data sets. The dot line represents the LBP histogram of the target-contained TDS image in the primary cell. The another line denotes the averaged LBP histogram on the 10 clutter-only TDS images. In the code 0, 3, 4, and 8, it shows a deviation between the target-contained histogram and the clutter-only histogram. 
}
    \label{fig.hist.comparison}
\end{figure}

\begin{figure}[!t]
    \includegraphics[width=\linewidth]{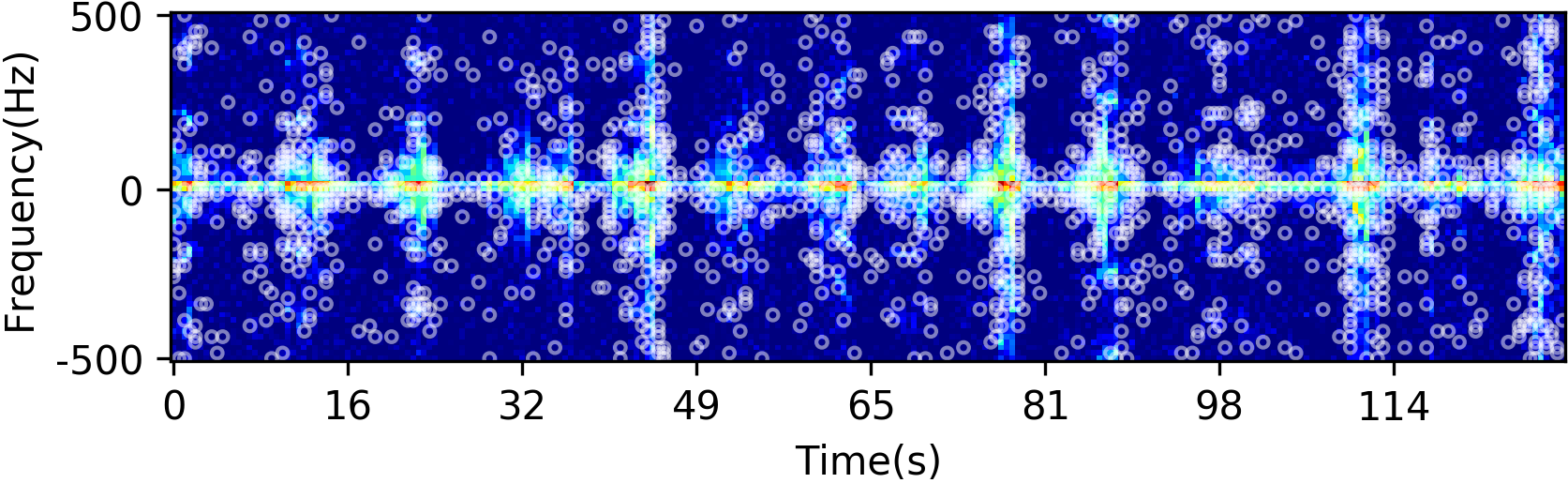}%
    \caption{Highlight the pixels which own 3 or 4 LBP code in the TDS image of the target-contained cell with white circles. These pixels locate in the contours of the blobs near the zero-frequency and on the edges of the speckles in the medium and high frequency. }
    \label{fig.tds.code34}
\end{figure}

In order to illustrate the numerical differences between the target-contained and the clutter-only in the LBP feature space, in this subsection, we compute the LBP histograms for these two type of TDS images in file indexed 135603 of Dartmouth(1993) data sets. In Fig.\ref{fig.hist.comparison}, the red dot line represents the target-contained LBP histogram, while the black line denotes the averaged histogram of all the clutter-only cells.  
In the tested file, there are only one primary cell and 10 clutter-only cells. Each cell contains about 2 minutes coherent returns. Using the method introduced in \ref{subsec.tds}, it obtains 11 TDS images and their LBP histograms. To reduce the number of the labels in Fig.\ref{fig.hist.comparison}, we average all the 10 clutter-only histograms to represent the clutter's characteristic. 

Let $\pmb{x}_0$ be the target-contained histogram in Fig.\ref{fig.hist.comparison} and $\pmb{x}_i, i \in [1,...,10]$ be the clutter-only histograms. Averaged clutter-only histogram is $\bar{\pmb{x}}=\sum\nolimits_{i=1}^{10}\pmb{x}_i/10$, and the standard deviation $\pmb{\sigma}_c$ is about
\begin{align}
 \pmb{\sigma}_c=&\left[0.004,0.002,0.001,0.003, 0.002,\right. \notag
 \\&\left. 0.001,0.001, 0.001, 0.01\right]^T, \notag
 \end{align}
 \textbf{which is close to the zero vector}. The deviation between the target-contained histogram and the averaged clutter-only histogram is denoted as $\pmb{\sigma}_t = |\pmb{x}_0 - \bar{\pmb{x}}|$. Now the target to clutter deviation ratio is termed as TCR:   
\beq\label{eq.tcr}
\textrm{TCR} = 10\log_{10}\frac{\pmb{\sigma}_t\cdot\pmb{\sigma}_t}{\pmb{\sigma}_c\cdot\pmb{\sigma}_c}. 
\eeq
Here the ` $\cdot$ ' operator means dot product operation.
The TCR in `HH' mode of the tested file in Fig.\ref{fig.hist.comparison} equals 12.5db, which is a \textbf{strong index} to show the separability between the target and clutter in the LBP histogram. In the Fig.\ref{fig.hist.comparison}, target-contained histogram gets more 3, 4 LBP codes  and less 0, 8 codes than the clutter-only. According to the definition of the`uniform' LBP,  code 3 and 4 denote the edges in the texture, code 0 and 8 mean the spot point and the flat zone respectively. \textbf{It implies that target-contained TDS image has more rough edges but less flat regions.} To visually testify this judgement, we draw the white circles directly on the TDS image of the target-contained primary cell, where the pixels get LBP code 3 or 4  in Fig.\ref{fig.tds.code34}. It is found that these circles are located on the contours of the blobs near the zero frequency band and at the edges of the speckles in the medium and the high frequency zone. These pixels are the very pixels which have varied texture compared to the clutter-only cells in the Fig.\ref{fig.tds.comparison}.

\section{Target detection via one-class SVM}
\label{sec.svm}
Assuming that  a big data set $D$ had the underlying probability distribution $p$, and a subset of $D$ named $S$ is observed, Sch\"{o}lkopf in  \cite{ScholkopfNC01} introduced one-class SVM to uncover that the probability of a test point drawn from $p$ lay outside of $S$ is bounded by a parameter $\nu \in (0,1)$. Later this method became known as $\nu$-SVM. 
In \cite{TaxML04}, $\nu$-SVM is proved to be able to find the spherical boundary of the intra-class samples in the Gaussian kernel space. Therefore we propose to use $\nu$-SVM to model the description of the sea clutter in the feature space of LBP, and view the sample outside the trained boundary of the clutter as an outlier. Here the outlier detection is equal to the target detection.
\subsection{Preliminary on $\nu$-SVM}
\label{subsec.presvm}
To begin a short description on $\nu$-SVM, we introduce some notations here. We denote the training samples from a subset $X$ as
\beq
\pmb{x}_1,...,\pmb{x}_m \in X,
\eeq
where the $m \in \mathbb{N}$ is the number of samples and the $\pmb{x}$ in bold font represents the feature vector of a sample, e.g. the histogram. Mapping the input vector into a dot product space $\mathbb{F}$ via function $\Phi$, it is found that the dot product of $\Phi$ can be evaluated by certain kernel functions \cite{Scholkopf02book} in the form
\beq
\label{eq.kf}
k\left(\pmb{x}, \pmb{y}\right) = \left(\Phi\left(\pmb{x}\right)\cdot  \Phi\left(\pmb{y}\right)\right),
\eeq
where the ` $\cdot$ ' operator means dot product operation.
The common Gaussian kernel is written as
\beq
\label{eq.kf.gk}
k\left(\pmb{x}, \pmb{y}\right) = \exp \left(-\frac{\|\pmb{x}-\pmb{y}\|^2}{s}\right),
\eeq
where $s$ is a positive scalar. 
Gaussian kernel holds two properties:
\beq
\label{eq.gk.norm}
k\left(\pmb{x}, \pmb{x}\right) = 1,
\eeq
and
\beq
\label{eq.gk.ieq}
k\left(\pmb{x}, \pmb{x}\right)=1 > k\left(\pmb{x}, \pmb{y}\right), \quad \textrm{for } \forall \pmb{x} \neq \pmb{y}.
\eeq

It is mentioned that if the input sample is mapped into the unit norm space (such as the Gaussian kernel space), training $\nu$-SVM is equal to find the minimum volume of the ball containing most of the intra-class samples in that space \cite{ScholkopfNC01, TaxML04}, for a $\nu \in (0,1)$, as:
\beq
\label{eq.vsvm.obj}
\underset{R\in \mathbb{R}, \pmb{\xi} \in \mathbb{R}^m, \pmb{c} \in \mathbb{F}}{min} \!\; \;  R^2+\frac{1}{\nu m}\sum\nolimits_{i=1}^{m}\xi_i,
\eeq

\beq
\label{eq.vsvm.cst}
\begin{split}
\textrm{subject to: } &\|\Phi\left(\pmb{x}_i\right)-\pmb{c}\|^2 \leqslant  R^2+\xi_i ,
\\ &\xi_i \geqslant0, i=1,...,m,
\\ &R \geqslant 0.
\end{split}
\eeq
Here the parameter $R$ and $\pmb{c}$ define the radius and center of the ball in the kernel space $\mathbb{F}$ respectively. $m$ indicates the number of training samples. None-negative slack variable $\xi_i$ accounts for the tolerated error. Parameter $\nu$ control the trade-off between the volume and accuracy in the training. Smaller $\nu$ means more penalty on the error of the classification $\xi_i$ (bigger $R$ is better), while bigger $\nu$ stresses more on the radius $R$ (smaller $R$ is better). 
To solve the constrained optimization, by using multipliers $\alpha_i, \beta_i \geqslant 0$, the objective function of $\nu$-SVM converts to the Lagrangian form:
\beq
\label{eq.vsvm.L}
\begin{split}
&L\left(R,\pmb{c}, \alpha_i, \beta_i, \xi_i\right)=R^2+\frac{1}{\nu m}\sum\nolimits_{i}\xi_i- 
\\& \sum\nolimits_{i} \alpha_i\left(R^2+\xi_i-\|\Phi\left(\pmb{x}_i\right)-\pmb{c}\|^2\right)-\sum\nolimits_{i}\beta_i\xi_i.
\end{split}
\eeq
 Then $L$ is minimized with respect to $R$, $\pmb{c}$ and $\xi_i$. Setting partial derivatives to zero, obtain:
\beq
\label{eq.vsvm.alpha}
\sum\nolimits_{i}\alpha_i=1, \quad \quad 0\leqslant \alpha_i \leqslant 1/\left(\nu m\right),
\eeq
and
\beq
\label{eq.vsvm.c}
\pmb{c}=\sum\nolimits_{i}\alpha_i\Phi\left(\pmb{x}_i\right).
\eeq
Substituting \eqref{eq.vsvm.alpha}  \eqref{eq.vsvm.c} 
into \eqref{eq.vsvm.L}, leads to the dual:
\beq
\label{eq.vsvm.dual}
\begin{split}
&\underset{\pmb{\alpha}}{min} \!\; \;  \sum\nolimits_{i=1}^{m} \sum\nolimits_{j=1}^{m}\alpha_i\alpha_j k\left(\pmb{x}_i, \pmb{x}_j\right)-\sum\nolimits_{i=1}^{m}\alpha_i k\left(\pmb{x}_i, \pmb{x}_i\right), 
\\
&
\textrm{subject to: } \quad 0\leqslant \alpha_i \leqslant 1/\left(\nu m\right), \sum \nolimits_{i} \alpha_i =1.
\end{split}
\eeq
Once the $\pmb{\alpha}$ is solved, the center $\pmb{c}$ of the spherical boundary is able to be computed by \eqref{eq.vsvm.c}. The radius $R$ equals $\|\Phi\left(\pmb{x}_k\right) - \pmb{c} \|^2$, where the $\Phi\left(\pmb{x}_k\right)$ is a vector on the boundary.
Now the decision function of the $\nu$-SVM for a testing sample $\pmb{x}$ is solved by:
\begin{align}
\label{eq.vsvm.df}
f\left(\pmb{x}\right)&=\textrm{sgn } \left( R^2 - \|\Phi\left(\pmb{x}\right)-\pmb{c}\|^2\right)
\\
&= \textrm{sgn } \left( R^2 - \sum\nolimits_{i,j}\alpha_i\alpha_jk\left(\pmb{x}_i, \pmb{x}_j\right)+ \right. \notag
\\& \qquad\qquad \left.2\sum\nolimits_{i}\alpha_ik\left(\pmb{x}_i, \pmb{x}\right)-k\left(\pmb{x}, \pmb{x}\right)\right).
\end{align}

If the $\pmb{x}$ locates outside of the ball, negative input of the sign function outputs $-1$. 

To solve the dual optimization of \eqref{eq.vsvm.dual}, conventional $\nu$-SVM needs a great amount of training samples in the \textbf{same} class to capture the real boundary.
However, TDS samples are sensitive to the weather conditions and sea states, we can not use the samples in one environmental condition to represent all the sea clutters. Therefore we only have limited number of TDS samples for the sea clutter in particular wind and sea states. 
Take the IPIX data set for example,  there is only one target-contained primary cell among the adjacent 10 or 20 plus clutter-only cells in each environmental condition. It is meaningless to train the $\nu$-SVM using all the clutter-only TDS images in one condition, which is hard to be generalized to other sea conditions, and test it only on the left one primary cell. \textbf{In practice, it is needed to train the $\nu$-SVM with both clutter and target samples for all the adjacent cells in each weather condition, and directly find out the target cell, with the  prior knowledge that there is only one outlier in the training samples.}
In the next subsection, we will prove that by setting a proper $\nu$ the outlier could be directly selected from the impure training samples.

\subsection{Classify the target by training $\nu$-SVM with impure clutter samples}
\label{subsec.imb.svm}

Conventional $\nu$-SVM model the boundary of one-class data with the pure positive samples.
This is not suitable for the TDS samples taken from the real Radar data, which is composed of a target-contained sample and a few of clutter samples in one weather condition. Because after training the one-class clutter samples in one environmental condition, the learned clutter boundary can not be generalized to other clutter caused by different wind and sea states. Furthermore, it is not feasible to label all the clutters 
in advance. 

Since in the real radar data sets, such as IPIX, the samples are impure (the majority is clutter samples but contains one target sample), intuitively the major samples would form a compact ball in the feature spaces. If the LBP feature of the primary cell are distinguished from the clutter-only cells, it would violate the stable ball constraint and get closer to the separating hyperplane. 
In this subsection, \textbf{we propose the proposition that the outlier is more closer to the boundary which is learned from the impure samples.} Based on this proposition, we design a framework to detect the target-contained TDS image via $\nu$-SVM classifier. 


\textbf{Proposition}:   Given a target-contained sample $\pmb{x}_0$ and $m-1$ clutter-only samples $\pmb{x}_i$, $i=1,...,m-1$, the boundary of the ball contains most of the samples in the Gaussian kernel space is more closer to the outlier $\pmb{x}_0$ than any clutter-only sample $\pmb{x}_i$:
\beq
\label{eq.ieqdist}
\|\Phi\left(\pmb{x}_0\right)-\pmb{c}\|^2 > \|\Phi\left(\pmb{x}_i\right)-\pmb{c}\|^2,
\quad\quad \textrm{for } \forall i \in [1,m-1].
\eeq
Here the $\pmb{c}$ is the center of the ball. Function $\Phi(\cdot)$ maps the feature vector to the Gaussian kernel space.

\textbf{Proof}: Expand the norm, \eqref{eq.ieqdist} changes to:
\begin{align}
\|\Phi\left(\pmb{x}_0\right)\|^2+\pmb{c}^2-2\Phi\left(\pmb{x}_0\right)\cdot \pmb{c} &> \|\Phi\left(\pmb{x}_i\right)\|^2+\pmb{c}^2-2\Phi\left(\pmb{x}_i\right)\cdot \pmb{c} 
\label{eq.ieqdot1}
\\
\Phi\left(\pmb{x}_0\right)\cdot \pmb{c}  &< \Phi\left(\pmb{x}_i\right)\cdot \pmb{c} 
\label{eq.ieqdot2}
\\
\left(\Phi\left(\pmb{x}_0\right) - \Phi\left(\pmb{x}_i\right) \right) \cdot\pmb{c} &< 0
\label{eq.ieqdot3}
\end{align}

The simplification from \eqref{eq.ieqdot1} to \eqref{eq.ieqdot2} is based on the Gaussian kernel property: $\|\Phi\left(\pmb{x}\right)\|^2 = k\left(\pmb{x},\pmb{x}\right)=1$.
From \eqref{eq.vsvm.c}, we know $\pmb{c}$ is the linear combination of $\Phi\left(\pmb{x}_i\right)$ and $\Phi\left(\pmb{x}_0\right)$. Substituting \eqref{eq.vsvm.c} into \eqref{eq.ieqdot3} leads to:
\beq
\label{eq.ieq.oi}
\sum\nolimits_{j=0}^{m-1} \alpha_j \left(\Phi\left(\pmb{x}_0\right) \cdot \Phi\left(\pmb{x}_j\right) - \Phi\left(\pmb{x}_i\right) \cdot \Phi\left(\pmb{x}_j\right) \right) < 0, \textrm{ for } \forall i \in [1,m-1].
\eeq

Put the kernel function of \eqref{eq.kf} into \eqref{eq.ieq.oi}, the proposition now becomes:
\beq\label{eq.dist.proposal}
\sum\nolimits_{j=0}^{m-1}\alpha_j\left(k\left(\pmb{x}_0, \pmb{x}_j\right)-k\left(\pmb{x}_i, \pmb{x}_j\right)\right) < 0, \textrm{ for } \forall i \in [1,m-1].
\eeq

Since in subsection \ref{subsec.LBPsepa} we observe that clutter-only samples $\pmb{x}_i, i\in[1,m-1]$ hold similar LBP histogram to the average $\bar{\pmb{x}}$ and the standard deviation $\pmb{\sigma}_c$ is close to the zero vector. Therefore the norm distance between the outlier $\pmb{x}_0$ and the clutter sample $\pmb{x}_i$ is approximated by:

\beq
\label{eq.dist.oave}
\begin{split}
\| \pmb{x}_0 - \pmb{x}_i \|^2 &\approx \| \pmb{x}_0 - \bar{\pmb{x}} - \pmb{\sigma}_c \|^2 \\
&= \| \pmb{x}_0 - \bar{\pmb{x}} \|^2 + \pmb{\sigma}_c\cdot\pmb{\sigma}_c -2\left(\pmb{x}_0  - \bar{\pmb{x}}\right) \cdot \pmb{\sigma}_c \\
&\approx \| \pmb{x}_0 - \bar{\pmb{x}} \|^2
\end{split}
\eeq
Substitute the  \eqref{eq.dist.oave} into the Gaussian kernel \eqref{eq.kf.gk}, it obtains:
\beq
\label{eq.kernel.io}
k\left(\pmb{x}_0, \pmb{x}_i\right) \approx k\left(\pmb{x}_0, \bar{\pmb{x}}\right)
\quad \textrm{for } \forall i \in [1, m-1].
\eeq
When both $\pmb{x}_i$ and $\pmb{x}_j$ are taken from the clutter-only samples. We have:
\beq \label{eq.dist.inner}
\begin{split}
\|\pmb{x}_j - \pmb{x}_i \|^2 &\approx \| \bar{\pmb{x}} - \pmb{\sigma}_c  - \bar{\pmb{x}} + \pmb{\sigma}_c\|^2 \\
&=  0\quad \textrm{for } \forall i,j \in [1, m-1].
\end{split}
\eeq

Therefore,
\beq
\label{eq.kernel.ii}
k\left(\pmb{x}_i, \pmb{x}_j\right) \approx 1, \quad
\textrm{for } \forall i,j \in [1, m-1].
\eeq

Now expand the left side of the \eqref{eq.dist.proposal}, we have

\begin{align}
&\sum\nolimits_{j}\alpha_j 
\left(k\left(\pmb{x}_0, \pmb{x}_j\right)-k\left(\pmb{x}_i, \pmb{x}_j\right)\right)
\notag
\\
&=\alpha_0 \left(k\left(\pmb{x}_0, \pmb{x}_0\right) - k\left(\pmb{x}_i, \pmb{x}_0\right)\right) + 
\alpha_1 \left(k\left(\pmb{x}_0, \pmb{x}_1\right) - k\left(\pmb{x}_i, \pmb{x}_1\right)\right) \notag
\\&\quad + ... +
\alpha_{m-1} \left(k\left(\pmb{x}_0, \pmb{x}_{m-1}\right) - k\left(\pmb{x}_i, \pmb{x}_{m-1}\right)\right)
\label{eq.dist.step1}
\\
&\approx \alpha_0 \left( 1 - k\left(\pmb{x}_0, \bar{\pmb{x}}\right)\right) +
  \alpha_1 \left(k\left(\pmb{x}_0, \bar{\pmb{x}}\right) - 1\right)  
  \notag
  \\&\quad
  + ... +\alpha_{m-1} \left(k\left(\pmb{x}_0, \bar{\pmb{x}}\right) - 1\right)  
\notag
\\
&= \left( \sum_{j=1}^{m-1}\alpha_j - \alpha_0\right)\left(k\left(\pmb{x}_0, \bar{\pmb{x}}\right) - 1\right) 
\label{eq.dist.step2}
\\
&= \left( 1 - 2\alpha_0\right)\left(k\left(\pmb{x}_0, \bar{\pmb{x}}\right) - 1\right)
\label{eq.dist.step3}
\end{align}

Put \eqref{eq.kernel.io} and \eqref{eq.kernel.ii} into \eqref{eq.dist.step1}, and use the constraint \eqref{eq.vsvm.alpha} in \eqref{eq.dist.step2}, it leads to \eqref{eq.dist.step3}.
If the outlier $\pmb{x}_0$, the LBP histogram of the target-contained TDS image, is not equal to the averaged LBP histogram of the clutter-only TDS images $\bar{\pmb{x}}$. According to \eqref{eq.gk.ieq}, it results $\left(k\left(\pmb{x}_0, \bar{\pmb{x}}\right) - 1\right)<0$.
%
Since the constraint of the objective function \eqref{eq.vsvm.dual} requires $0\leqslant \alpha_i \leqslant 1/\left(\nu m\right)$, setting the $\nu > 2/\left(m\right)$ would make $\alpha_0<1/2$ for sure. In this condition, \eqref{eq.dist.step3} is less than 0, our proposition \eqref{eq.dist.proposal} is valid. \hfill\(\blacksquare\)

Since we have proved that the outlier is more closer to the boundary of the impure training samples of $\nu$-SVM, in the following, we state our target-detection procedure in four steps:

Step 1: Convert the time-sequential returns, which are collecting from both $m-1$ clutter-only cells and 1 target-contained cell, to $m$ TDS images.

Step 2: Compute the LBP histogram $\pmb{x}_j, j\in[0,m-1]$, of the $m$ images and store them in a training set $X$.

Step 3: Train the $\nu$-SVM with $\nu > 2/(m)$ in $X$ and learn the parameters for the decision function $f$ in \eqref{eq.vsvm.df}.

Step 4: Compute the distances to the boundary for each sample (This is the input of the decision function $f$). The one with least distance is classified as the target-contained.

\section{Experiments}
\label{sec.exp}


In this section, to test the proposed method, we have chosen the same twenty IPIX data sets, which are used in the \cite{HuTAP06, LuoGRSL13, ShiTGRS18, ShuiTAES14}. We first list the parameters used for preparing TDS image, for computing LBP histogram and for training of $\nu$-SVM. The experimental results are discussed after.

In our experiment, we divide the $2^{17}$-length sequential returns of the IPIX 1993 data set into $512$-length non-overlapped segments. Each segment is re-scaled to 64 after the hamming-windowed Fourier transformation. The resolution for a TDS image is $w=256, h=64$ in 1993 data sets. For the $60000$-length data in 1998 data sets, we use the $256$-length segments. Now the width equals to $w=234$, which is close to $256$. The final resolution of a TDS image in 1998 data sets is $w=234, h=64$. 

To compute the LBP code, we choose $P=8$ circular neighbour pixels with radius $R=1$ around the central pixel. Set the LBP in `uniform' format and define the LBP histogram in 9 bins. To train the $\nu$-SVM, we choose the Gaussian kernel with bandwidth $s = 1/m$. Here $m$ is the number of samples, which equals 10+ and 20+ in 1993 and 1999 data sets respectively.
To satisfy the proposition \eqref{eq.dist.proposal} for the training with impure samples, the key parameter $\nu$ is set to 0.4, which is bigger than $2/m$ in both data sets.  
\begin{table}[!ht]
\centering
\caption{TCR and detection results of 4 polarized modes in IPIX.}
\label{tab.tcr}
\begin{tabular}{l c c c c c c}
\hline\hline
\tabincell{l}{Dartmouth(1993) \\ Index No.} & \tabincell{c}{TCR\_HH\\(db)}     & \tabincell{c}{TCR\_HV\\(db)}        & \tabincell{c}{TCR\_VH\\(db)}    & \tabincell{c}{TCR\_VV\\(db)}  \\
\hline
135603              & 12.5(\cmark)    & 0.19(\xmark)       & 12.2(\cmark)   & 8.68(\cmark) \\
220902              & 6.76(\cmark)    & 11.9(\cmark)       & 4.08(\cmark)   & 3.81(\xmark) \\
191449              & -1.3(\xmark)    & 3.01(\cmark)       & 0.51(\xmark)   & -2.3(\xmark) \\
202217              & 11.7(\cmark)    & \textbf{5.27}(\xmark)       & 4.50(\cmark)   & 11.7(\cmark) \\
001635              & 15.1(\cmark)    & 11.0(\cmark)       & 7.25(\cmark)   & 9.40(\cmark) \\
163625              & 12.1(\cmark)    & 9.56(\cmark)       & 11.4(\cmark)   & 4.75(\cmark) \\
023604              & 13.6(\cmark)    & 3.01(\xmark)       & 12.1(\cmark)   & 8.04(\cmark) \\
162155              & 8.02(\cmark)    & 16.6(\cmark)       & 11.8(\cmark)   & 0.99(\xmark) \\
162658              & 13.2(\cmark)    & 17.6(\cmark)       & 18.7(\cmark)   & 8.89(\cmark) \\
174259              & 9.82(\cmark)    & 9.96(\cmark)       & 11.2(\cmark)   & 7.11(\cmark) \\
\hline
\tabincell{l}{Grimsby(1998) \\ Index No.} & \tabincell{c}{TCR\_HH\\(db)}     & \tabincell{c}{TCR\_HV\\(db)}        & \tabincell{c}{TCR\_VH\\(db)}    & \tabincell{c}{TCR\_VV\\(db)}  \\
\hline
163113              & -9.2(\xmark)    & 11.0(\cmark)       & 8.87(\cmark)   & -3.8(\xmark) \\
202225              & 10.2(\cmark)    & 15.3(\cmark)       & 15.6(\cmark)   & 9.96(\cmark) \\
202525              & 9.67(\cmark)    & 15.9(\cmark)       & 15.2(\cmark)   & 9.49(\cmark) \\
171437              & 10.8(\cmark)    & 17.1(\cmark)       & 14.5(\cmark)   & 9.88(\cmark) \\
180588              & 13.8(\cmark)    & 17.9(\cmark)       & 14.8(\cmark)   & 13.0(\cmark) \\
195704              & 9.12(\cmark)    & 16.5(\cmark)       & 12.9(\cmark)   & 8.67(\cmark) \\
164055              & \textbf{7.42}(\xmark)    & 13.6(\cmark)       & 11.6(\cmark)   & 5.28(\xmark) \\
173317              & 8.26(\cmark)    & 13.1(\cmark)       & 14.8(\cmark)   & 11.1(\cmark) \\
173950              & 11.4(\cmark)    & 15.5(\cmark)       & 15.8(\cmark)   & 12.3(\cmark) \\
184537              & 5.61(\xmark)    & -1.3(\xmark)       & 1.13(\xmark)   & 3.46(\xmark) \\
\hline\hline
\end{tabular}
\end{table}

Table \ref{tab.tcr} records all the TCR and detection results in the tested 20 files in IPIX. First column lists the file index. Latter four columns show the target to clutter ratio (TCR) \eqref{eq.tcr} with respect to the deviation to the averaged clutter in four modes.  The `\cmark' in the brackets denotes correct detection in the final test, symbol`\xmark' means wrong.
It shows that the higher TCR brings better detection performance. This also proves that the separable features have a great influence on the detection results of the classifier.   It is noted that when TCR is under \textbf{5.27} db in Dartmouth(1993) data sets, the detection results is not reliable. In Crimsby(1998) it raises to \textbf{7.42} db. This is because the clutter-only cells in Crimsby(1998) is twice times more than the cells in Dartmouth(1993).

\begin{table}[!ht]
\centering
\caption{Comparisons of detection rate}
\label{tab.dr}
\begin{tabular}{l c c c c c c}
\hline\hline
Dartmouth(1993)     & Our             & NTFD\cite{ShiTGRS18}    & Tri\cite{ShuiTAES14} & Factral\cite{HuTAP06, LuoGRSL13}  \\
HH                  & \textbf{0.90}   &  0.754         & 0.577      & 0.303    \\
HV                  &  0.70           &\textbf{0.761}  & 0.661      & 0.468    \\  
VH                  & \textbf{0.90}   & 0.75           & 0.65       & 0.453    \\
VV                  & \textbf{0.70}   & 0.672          & 0.543      & 0.387    \\
\hline
Grimsby(1998)       & Our             & NTFD\cite{ShiTGRS18}    & Tri\cite{ShuiTAES14} & Factral\cite{HuTAP06, LuoGRSL13}  \\
HH                  & 0.70            &\textbf{0.903}& 0.65         & 0.292    \\
HV                  & 0.90            &\textbf{0.997}  & 0.934      & 0.604    \\  
VH                  & 0.90            & \textbf{1}     & 0.967      & 0.687    \\
VV                  & 0.70            & \textbf{0.878} & 0.658      & 0.283    \\
\hline
Total               & Our             & NTFD\cite{ShiTGRS18}    & Tri\cite{ShuiTAES14} & Factral\cite{HuTAP06, LuoGRSL13}  \\
HH                  & 0.80            &\textbf{0.82}   & 0.62       & 0.298    \\
HV                  & 0.80            &\textbf{0.88}   & 0.8        & 0.536    \\  
VH                  & \textbf{0.90}   & 0.88           & 0.81       & 0.570    \\
VV                  & 0.70            & \textbf{0.79}  & 0.6        & 0.335    \\
\hline\hline
\end{tabular}
\end{table}
To validate the proposed detector, we have compared it with other three methods. Table \ref{tab.dr} shows the detection rate of the twenty files from 1993 and 1998 data sets. The rate in bold font means it ranks first in the four methods. It shows that our method get advanced results in the 1993 data. Detection rate is high on the `HV' and `VH' model in 1998 data. In all the twenty data, LBP feature holds comparable performance to the NTFD which fuses multiple features. It proves that LBP could be fused as a general feature for further improving the detection rate.

\section{Conclusion}
\label{sec.con}
In this paper, we model the texture of TDS image in LBP histogram. Based on the one-class SVM classifier, we interpret the outlier of the detector as the target, whose appearance causes the interference of the underlying motion of waves in the time-frequency domain. In our experiment tested on the IPIX data sets, we have noted that the existing detectors can not detect the small-floating target perfectly in all the environmental conditions. The pursuit of more distinguished features is still the trend of  detecting small target in the clutter. We believe that more advanced visual feature which can describe the changes of the sea clutter in more details will have a great potential.

\bibliographystyle{IEEEtran}
\bibliography{./bib/jdat2020.bib}


\end{document}